# The many facets of academic mobility and its impact on scholars' career

Fakhri Momeni[1], Fariba Karimi[2], Philipp Mayr[1], Isabella Peters[3] and Stefan Dietze[1,4]

[1] *firstname.lastname@gesis.org*
GESIS – Leibniz Institute for the Social Sciences, Unter Sachsenhausen 6-8, 50667 Cologne (Germany)

[2] *karimi@csh.ac.at*
Complexity Science Hub Vienna, Josefstädter Straße 39,1080 Vienna (Austria)

[3] *I.Peters@zbw.eu*
ZBW – Leibniz Information Centre for Economics, Düsternbrooker Weg 120, 24105 Kiel (Germany)

[4] *stefan.dietze@hhu.de*
Heinrich-Heine-University Düsseldorf, Universitätsstr. 1, 40225 Düsseldorf (Germany)

**Corresponding author:**
Fakhri Momeni
GESIS -- Leibniz Institute for the Social Science
Unter Sachsenhausen 6-8
50667 Köln
Tel: 0221 47694-544
Mail: fakhri.momeni@gesis.org

**Highlights:**

- We use Scopus data to identify and explore multiple facets of academic mobility among researchers from various fields and countries in different career stages.
- We infer the gender of scholars that enables comparative analyses between two genders.
- The mobility of scholars is tracked over time using scholarly publications and considering the frequency of movements in analyses that leads to more comprehensive results.
- We investigate the impact of mobility on scholars' publications and citations.
- We investigate the effect of mobility on large-scale collaboration networks to display the position and role of mobile researchers in these networks.

Abstract

International mobility in academia can enhance the human and social capital of researchers and consequently their scientific outcome. However, there is still a very limited understanding of the different mobility patterns among scholars with various socio-demographic characteristics. By studying these differences, we can detect inequalities in access to scholarly networks across borders, which can cause disparities in scientific advancement. The aim of this study is twofold. First, we investigate to what extent individuals' factors (e.g., country, career stage, and field of research) associate with the mobility of male and female researchers. Second, we explore the relationship between mobility and scientific activity and impact. For this purpose, we used a bibliometric approach to track the mobility of authors. To compare the researchers' scientific outcomes, we considered the number of publications and received citations as indicators, as well as the number of unique co-authors in all their publications. We also analysed the co-authorship network of researchers and compared centrality measures of "mobile" and "non-mobile" researchers. Results show that researchers from North America and Sub-Saharan Africa, particularly female ones, have the lowest, respectively, highest tendency towards international mobility. Having international co-authors increases the probability of international movement. Our findings uncover gender inequality in international mobility across scientific fields and countries. Across genders, researchers in the Physical sciences have the most and in the Social sciences the least rate of mobility. We observed more mobility for Social scientists at the advanced career stage, while researchers in other fields prefer to move at earlier career stages. Also, we found a positive correlation between mobility and scientific outcomes, but no apparent difference between females and males. Indeed, researchers who have started mobility at the advanced career stages had a better scientific outcome. Comparing the centrality of mobile and non-mobile researchers in the co-authorship networks reveals a higher social capital advantage for mobile researchers.

**Keywords:** International academic mobility, gender inequality, co-authorship network, bibliometrics, scientific performance, scientific success

1. **Introduction**

Scientific progress is the result of a collaborative process that involves researchers across the world and international collaboration. For that, mobility of researchers is important, since it fosters communication, collaboration and knowledge transfer between researchers – all factors which are considered crucial for scientific progress as well as for success and performance of researchers. However, research has shown that extent and distribution of mobility is gender-dependent (El-Ouahi et al., 2021; Jöns, 2011; Leemann, 2010; Ryazanova & McNamara, 2019). For example, although participation in science of researchers that have been identified as females has seen a significant advancement during the last years (Carr et al., 2015; Ovseiko et al., 2017), the vast majority of female researchers are not mobile (Jöns, 2011).

Gender inequality in academic international mobility needs to be tackled in many societies, because of the existing troubles to go abroad for women. Depending on the family situations during the career life, women can face more barriers to being mobile at any career stage (Jöns, 2011). Identifying differences in mobility among various countries, scientific fields and the career stages can help to better explore the root of problems women deal with.

Studying the mobility patterns among different societies enhances our understanding of the researchers' motivations and restrictions to move internationally. Someone moves to a new country due to experience working with colleagues and researchers in another environment, while another one moves away from problems in the current country (e.g., gender inequality, lack of labor in a particular field, financial and political grounds, etc.), which may as well be a potential barrier to going aboard. On the other hand, the chance of obtaining an appropriate research position abroad is not the same for all groups, which affects the decision of researchers and mobility directions. Some fields are more in demand in a particular country and academic positions are offered to international or inexperienced researchers regardless of gender, whereas, in some other fields or countries, positions are restricted to skilled or male researchers. In addition, the comparative analyses of mobility effects on researchers across various groups and different career phases, reveal the importance of mobility between countries. By applying a bibliometric approach, we can track the mobility of researchers through their publications. Some studies have already investigated international academic mobility in a similar way (Aman, 2018a; El-Ouahi et al., 2021; Petersen, 2018; Robinson-Garcia et al., 2019; Subbotin & Aref, 2021). *Table 1* summarizes the notable related studies in terms of academic mobility and the features they considered as well as the scope of analysed data. No prior work exists that investigates the role of gender in the context of mobility and scientific impact on a global scale. The main contribution of this paper is to comparatively study the mobility pattern of different genders, i.e., women and men, and their scientific outcome among scientific fields and countries across the stages of career development. This will reveal the extent of gender inequality in different societies. We define international academic mobility as changes in the scholars' country of affiliation over time. In the following, *'mobility'* refers to '*international academic mobility*'.

In this paper, we aim to answer two main research questions:

1. To what extent individuals' factors (e.g., country, career stage, and field of research) do associate with the mobility of researchers and how do they differ for males and females?

2. How do different characteristics of mobile researchers correlate with scientific outcomes of researchers?

For the first research question we investigate the role of gender, scientific field, country of origin, and international collaboration on the likelihood of becoming mobile and compare the mobility pattern of two genders, i.e., male and female, across countries and scientific fields and through three career stages. For the second research question, we use the number of publications, received citations and number of unique co-authors of researchers and examine the relationship between mobility and these indicators. We also analyse the co-authorship networks to present the differences between centrality measures of mobile and non-mobile researchers.

Our study shows the following novel aspects; firstly, the scale and broad coverage of the used dataset from various fields and countries, that contributes to the generalisability of the results. Second, inferring the gender of scholars with high accuracy on scale that enables comparative analyses between two genders. Third, tracking the mobility of scholars over time using scholarly publications and considering the frequency of movements in analyses that leads to more comprehensive results. Lastly, by applying centrality measures in large-scale collaboration networks at the individual level, we compare the position and role of mobile with non-mobile researchers in these networks.

## 2. Related work

In the academic world, communication and collaboration between scientists are crucial to individual and scientific success. International mobility can connect scholars from different countries with various scientific backgrounds and along with it may enhance knowledge exchange. It associates with both human capital (refers to the knowledge and experience of individuals (De Cleyn et al., 2015)) and social capital (as the resources available to individuals and groups through membership in social networks (Villalonga-Olives & Kawachi, 2015)) of researchers. Thus, it can affect the researchers' scientific impact positively by sharing, exchanging knowledge and obtaining other opportunities to enhance individual's skills or negatively by disconnecting from local co-authors and having difficulties making connections with new colleagues because of different languages and cultural backgrounds (Almansour, 2015; Caniglia et al., 2017). Also, at the country level, a researcher with the experience of staying abroad can act as a hub which mediates between different countries and, overall, increases collaboration between both countries (brain circulation) (Saxenian, 2007). However poor countries suffer from losing talents who migrate and labor shortage (brain drain) (Arrieta et al., 2017).

### 2.1. Mobility and field of research

Epistemic characteristics of fields influence decisions of researchers for national or international mobility (Laudel & Bielick, 2019). In some fields, human capital of researchers is more transferable, but some others are more specific for a country. For example, Bäker (2015) reports that researchers in disciplines with both quantitative and qualitative research methods (pluralistic) are more likely to lose their human capital after changing affiliation, because of the diversity in research approaches. Aman (2020) measured the knowledge transmission among mobile and non-mobile researchers and discovered highest knowledge transmission in "Earth and Planetary Sciences" and "Neurosciences". Depending on the size and domain of used data, prior studies have mentioned different proportions of mobility across disciplines. Cañibano et al. (2011) analysed a set of 10,000 PhD holders in Spain from *Scientific Information System of Andalusia* dataset (SICA) and reported the most international mobility in "social sciences" and "science and technology of health" as the least mobile discipline. In contrast, Subbotin and Aref (2021) found that Russian scientists have the most and least international mobility in Physical Sciences and Social Sciences, respectively.

### 2.2. Mobility and Gender

Gender inequality in science is more obvious in mobility, due to barriers for women to go abroad. It can decrease their visibility and scientific impact, as Kong et al. (2021) explained that women suffer from citation inequality due to first-mover advantage of men. However, Bozeman and Gaughan (2011) found no evidence that men or women adopt a "nationalist" strategy (wishing collaborators from one's own nation or shared language) in collaboration. Prior studies have shown that women are less likely to have international mobility (El-Ouahi et al., 2021; Jöns, 2011; Leemann, 2010; Ryazanova & McNamara, 2019). This varies by many factors such as discipline, career stage and country of origin. Bhandari (2017) showed a lower percentage of internationally mobile female researchers in STEM disciplines and the results of a study by Jöns (2011) report a less international mobility of women in natural sciences. Jayachandran (2015) showed that many poor countries favor men in mobility than women due to cultural norms. There are some mobility programs around the world that prioritize women. For example, women are over-represented by Erasmus mobility program (Böttcher et al., 2016; De Benedictis & Leoni, 2020). Jöns (2011) found that at the earlier career stages, male and female students are equally internationally mobile, but at advanced career stages flexibility of women to go abroad decreases much more than their male colleagues. However, Leemann (2010) revealed that the probability to move abroad decreases with age for both genders.

### 2.3. Mobility and academic impact

Academic mobility influences the co-authorship pattern that impacts quality and quantity of scientific productivity. These effects differ between disciplines with varying characteristics. Halevi et al. (2016) analysed the data of 100 top authors in seven disciplines and showed that for some disciplines, country mobility has a negative effect on productivity and received citations, while for others has a positive or no effect. Bäker (2015) analysed the impact of changing affiliation for economics as a less pluralistic discipline (e.g., only quantitative research methods) and management as a more pluralistic discipline (e.g., quantitative and qualitative research methods) and report a worse effect for the most pluralistic disciplines in the short-term, because researchers in those disciplines are more likely to lose human capital due to variety of approaches in the new institutions. Petersen (2018) reported an increase in co-author diversity as the effect of mobility for physics scientists. Wang et al. (2019) examined the change in collaboration patterns of mobile researchers and found an increase in domestic collaboration but at the cost of decreasing international collaboration. Also, Bernard et al. (2021) showed the reduced likelihood of collaboration with previous co-authors after mobility.

Also, the time of moving is a significant factor influencing academic outcomes. Zhao et al. (2020) found that the productivity of researchers who move to China at an earlier career stage is higher than those who move at a later stage. However, Bauder (2020) reports that mobility can lead to the loss of national social capital that negatively affects early-career researchers in particular. Furthermore, the results of a study by Ryazanova and McNamara (2019) indicate a negative effect of international mobility at the first postdoctoral researcher (postdoc) job on research productivity, however they found that an international movement between year 2 and 7 of a postdoc is better than later.

*2.4. Approaches, data resources and investigated features*

Many studies utilized qualitative approaches to this problem. The major drawback in qualitative analysis is mainly the small size of data as well as bias problems (Bäker, 2015; Bauder et al., 2017; Bedenlier, 2018; Cohen et al., 2020; Laudel & Bielick, 2019; Leung, 2017; Morano-Foadi, 2005; Nikunen & Lempiäinen, 2020; Schaer et al., 2017). Other studies with quantitative approaches employed resources such as CV (Cañibano et al., 2008; Laudel & Bielick, 2019; Li & Tang, 2019; Youtie et al., 2013; Zhao et al., 2020) and bibliometric data of researchers (Aman, 2018a; Chinchilla-Rodríguez et al., 2018; El-Ouahi et al., 2021; Petersen, 2018; Robinson-Garcia et al., 2019; Subbotin & Aref, 2021) to track their movements with larger data sample sizes. Table 1 shows these studies with a bibliometric approach and the investigated features as well as the set of selected authors. The most of these studies used a restricted set of authors or features. For example, Petersen (2018) analysed the *American Physical Society* (APS) dataset which covers publications in the domain of physics. Subbotin and Aref (2021) employed Scopus dataset for analysing the international migration of researchers who have published with Russian affiliation address, by discipline. El-Ouahi et al. (2021) investigated the international mobility for countries in the Middle East and North Africa region from Web of Science dataset. In this study they compared the Gender ratio of migrants for the countries in this region. Among all these, only the study by Robinson-Garcia et al (2019) covered all authors in Scopus from various countries and classified mobile authors into three groups (migrant, directional travelers and non-directional travelers). with those of non-mobile authors. Our study includes the authors from different countries too, but we rank the mobility of authors according to the frequency of changing their affiliated countries, which leads to a more detailed analysis of the extent of mobility. Gender, field, career stage and network centralities of researchers are other distinctive aspects of our study that enable us to discover the disparities and issues in mobility in different societies and scientific communities.

Table 1 Investigated features in studies that used a bibliometric approach to study international mobility of authors.

| Study | Investigated feature | | | | Restriction in author selection |
|---|---|---|---|---|---|
| | **Gender** | **Field** | **Country** | **Career stage** | |
| (Aman, 2018a) | ✗ | ✓ | ✗ | ✗ | Authors with German affiliation |
| (Petersen, 2018) | ✗ | ✗ | ✓ | ✗ | Physics researchers |
| (Robinson-Garcia et al., 2019) | ✗ | ✗ | ✓ | ✗ | - |
| (Subbotin & Aref, 2021) | ✓ | ✓ | ✗ | ✗ | Russian Authors |
| (El-Ouahi et al., 2021) | ✓ | ✗ | ✓ | ✓ | Authors with an affiliation in Middle East and North Africa region |
| Our study | ✓ | ✓ | ✓ | ✓ | - |

## 3. Data and methods

### 3.1. Data sources

The in-house Scopus database maintained by the German Competence Centre for Bibliometrics (Scopus-KB), 2020 version, is used as the main resource of analyses. We utilized publications indexed in Scopus to study the international mobility of scholars. In order to identify authors, we used Scopus author ID which enable us to track the international mobility of authors (Aman, 2018b). Kawashima and Tomizawa (2015) estimated the accuracy of Scopus author Id using KAKEN database (largest funding database in Japan) and found a very high precision (99%) and recall (98%).

For detecting the gender status, we apply a combined name and image-based approach introduced by Karimi et al. (2016). They tested the accuracy of this method in their paper with a sample of 693 male and 723 female names. The ground truth consists of a manually labelled random sample of academics, their full names, institutions, countries, and their gender. This method (combination of first names, family names, and images) has a general f-score of 93% which is higher than other existing gender inference methods and is more robust for different nationalities. The only exception is for Asian names, especially Chinese names, where this method has low accuracy. Therefore, we try to eliminate those ambiguous names to increase the accuracy of results for these countries. From 32,110,580 identifiers in Scopus, 7,956,823 had no first name or just initial among their publications that any gender detection methods would not be able to infer genders. For the remaining 24,153,757 identifiers, our gender inference method was able to infer the gender of 8,592,307 (~35%) names that could be identified.

We acknowledge that gender is a non-binary identity. For our purposes, and due to the lack of more fine-grained gender information, we consider it as binary in this work. The term "gender" doesn't refer to the sex of the authors, nor the gender that the authors identify themselves with. We refer to gender as the general societal convention in assigning first names to individuals in combination with what machine learning face recognition algorithms identify as female or male (Karimi et al., 2016). Hence, this work can only be a starting point for more detailed analyses of the role of gender on the mobility of researchers.

To detect the disciplines of authors we used the "All Science Journal Classification" (ASJC) system of Scopus[1] which contains 27 subject categories. Next, we classify these disciplines to four main fields according to the Scopus classification.

---

[1] More information: https://service.elsevier.com/app/answers/detail/a_id/14882/supporthub/scopus/~/what-are-the-most-frequent-subject-area-categories-and-classifications-used-in/ Accessed 14 Sep. 2021

The field with most publications is considered as the *main field* of the author. About 1.5% of the authors had more than one most popular field and we excluded these authors from the analyses.

*3.2. Career stages*

Several approaches to stratify researchers into career stages have been proposed and discussed. The major challenge lies in the individual situations of career progression, which is highly dependent on many factors (e.g., discipline, faculty, and career interruption). Although not optimal, this is why most approaches work with fixed time periods for every career stage. Li et al. (2019), for example, defined researchers as 'junior' within three years after their first publication. In contrast, Bäker (2015) recommended 4-6 years for this phase of career.

We agree with Bazeley (2003) and Bosanquet et al. (2017) who considered a period of five years as the minimum for the early career stage and therefore also adopted the approach presented by Mascarenhas et al. (2017) who used the following calculations for the three career stages:

- *Early career stage* ($s_1$): years from the first publication year to 4 years after it
- *Mid-career stage* ($s_2$): years between 5 and 9 years after the first publication year
- *Late career stage* ($s_3$): years more than 10 years after the first publication year

We did not find a clear definition of advanced career stages (middle and late) in other studies. The only study from Ponjuan et al. (2011) defined 4 to 5 years for mid-career and more than five years for late career stages for pre-tenure faculty members, which is close to our thresholds for these two career phases.

To further increase comparability and homogeneity among the studied researchers (and to disregard career paths that show too much variance) as mentioned above, we excluded authors who published less than one publication per three years.

*3.3. Mobility detection*

In Scopus, affiliation information as well as the country of affiliation of authors are separately available for each publication. We utilized the country information of the affiliation to track the mobility of authors. Since affiliation of the authors are provided for each publication, we can track the changes of affiliations over time.

Mobility in this study is defined as having a co-affiliation (affiliated with more than one country in the same publication) or multiple affiliations (affiliated with at least two countries in two papers) (Chinchilla-Rodríguez et al., 2017; Petersen, 2018). Therefore, an author with one affiliation country through the authors' publications is considered as non-mobile.

The *origin country* of the author is the country of author's affiliation on the first publication.

*Mobility score* is applied to measure the frequency of mobility. To calculate the mobility score of the author, we sort the lists of affiliation countries based on publishing years. Next, we compare the affiliation countries for each year to those from previous publishing year and assign one score for each country in the current year that doesn't exist in the list of previous year. Then, the sum of scores across all publishing years will be assigned as the mobility score of an author. For the first publishing year with non-empty list of countries, the number of unique countries except the first country which is the origin country, will be considered as the score for that year. Table 2 shows two examples of calculating the mobility score.

Table 2. Two examples calculating the mobility score. Author A has a mobility score of 5 and author B has a mobility score of 4.

|  | Author A |  |  | Author B |  |
|---|---|---|---|---|---|
| **Publishing year** | **Affiliation country** | **score** | **Publishing year** | **Affiliation country** | **score** |
| 2002 | USA | 0 | 2000 | Japan | 0 |
|  |  |  |  | China | 1 |
|  |  |  |  | Singapore | 1 |
| 2004 | Germany | 1 | 2005 | Japan | 0 |
|  | Canada | 1 |  | China | 0 |
|  | France | 1 |  |  |  |
| 2007 | USA | 1 | 2006 | Australia | 1 |
|  | Canada | 0 |  | Japan | 0 |
| 2008 | USA | 0 | 2008 | China | 1 |
|  | Germany | 1 |  |  |  |
| **Sum of scores** |  | 5 |  |  | 4 |

We select those authors for our analyses who have a Scopus author ID, gender status (male or female), and at least early and mid-career stage publications (10 years career age). We consider active authors who published at least one-third of their career age (e.g., an author having the first publication in 2001 and last publication in 2013 has a career age of 12 and should have at least four publications). To count the number of received citations, we apply a three years citation window after the publication year. To ensure that we count the received citations equally for all publications, we include all publications until 2016 and assume it as the last publication year for those authors published after this year. In addition, to have more authors with the late-career stage, we include the authors with the first publication year until 2002. Since most authors have their first publication from 1996, we exclude the authors with the first year before this year. By applying all these filters, we extract a list of 1,184,355 authors.

### 3.4. Region and Income level of countries

We use annual Gross Domestic Product (*GDP) per capita*, Purchasing Power Parity (PPP) (current international $) and region of countries from the World Bank[2] in the analyses. The average GDP per capita from 1996 to 2016 is considered for each country.

### 3.5. Mobility outcome

Similar to Barabási and Musciotto (2019) we define two concepts to assess the impact of mobility on the research outcome: performance and success. According to this definition "Performance is about individual effort, while success is a collective quantity capturing community's acknowledgment of effort and performance". Therefore, we evaluate the performance by *mean publication per year* (*PPY*) (by dividing the number of publications by career age) and success by *mean citation per publication* (*CPP*) (by dividing the sum of citations by number of publications). To calculate CPP, we consider the citations received until three years after the publication year (to account for differences in the age of publications). Finally, we use *Mean unique co-authors per publication* (*COPP*) (number of unique co-authors among all publications divided by number of publications) to measure the impact of mobility on co-author diversity.

While first/corresponding authors are considered to have made the major contribution to the paper, all publications of researchers would affect their career through the accumulation of citations and h-index. On the other hand, most researchers have their first-authored papers at the earlier years of their career life and their position in publications moves from first to the last author while progressing through career stages (Gingras et al., 2008; Way et al., 2017). Therefore, for calculating PPY, CPP, and COPP, we consider all publications that an author has co-authored and don't differentiate between various authorship positions.

---
[2] https://data.worldbank.org/indicator/NY.GDP.MKTP.PP.CD Accessed 14 Sep. 2021

### 3.6. Co-authorship network

We measure the social capital of authors by analysing the co-authorship network. We generate the co-authorship network for each discipline based on the publication records. In this network, every author is one node, and each edge represents a co-authorship activity between two authors. The number of shared publications between authors indicates the weight of the edges. The structure of the network changes and evolves over the years. Because of that, for each discipline and year, we create a network. We assume any two authors (nodes) are connected in a given year if they have at least one co-authorship in the past five years. Therefore, those nodes and edges related to publications older than five years are not included in the network of that year. Also, nodes without any connection are removed from the networks. The network analyses are performed with the *igraph*[3] library in Python.

Degree, closeness and betweenness are three well-known centrality measures employed in most previous works (Abbasi et al., 2011; Chinchilla-Rodríguez et al., 2017; Karimi et al., 2019; Li et al., 2013; Servia-Rodríguez et al., 2015) that analysed a co-authorship network. We use degree and closeness, and disregard betweenness because of its high calculation complexity and time expenditure for such large-scale collaboration networks on the author level.

Co-authorship networks consist of many communities in which co-authors are connected via their publications. *Clustering coefficient* is another centrality metric utilized in this study to observe how the collaboration patterns in communities modify when authors change their communities via mobility, as it was used by Abbasi et al. (2011).

Collaborating with top researchers is a motivation for many researchers to go abroad. We apply *coreness* centrality to examine whether they have better access to top authors.

We calculate these centrality measures for any node (author) in a given network with N nodes and E edges:

- *Degree:* Number of ties that the node has with other nodes (Freeman, 1978):

$$d_i = \sum_j a_{ij}$$

    where $a_{ij}$ is an element of the adjacency matrix and indicates the existence or non-existence of a link between node$i$ and node$j$.

- *Closeness*: Average length of the shortest path between the node and all other nodes in the graph (Freeman, 1978). It indicates how close an author is to all other authors in the network. Because this centrality is a global centrality and it is affected by the number of nodes in the network, we use the normalized closeness by multiplying raw closeness by total number of nodes except the one $(n - 1)$ (Mohammadamin et al., 2017):

$$C_i = \frac{n-1}{\sum_j e_{ij}}$$

    where $n$ is the number of nodes and $e_{ij}$ is the number of links in the shortest path from node $i$ to node $j$.

---

[3] https://igraph.org/python/ Accessed 14.9.2021

- *Coreness*: It represents how well a node is connected to other important nodes and also with periphery nodes in the network (Saxena & Iyengar, 2020). A k-core of $k$ indicates that a node is connected to a subset of nodes that have at least $k$ degree or higher.
- *Clustering coefficient:* The probability that the adjacent nodes of a node are connected. In co-authorship network, clustering indicates how likely it is that two co-authors of a given author are also co-authors (Zare-Farashbandi et al., 2014).

## 4. Results

First, we will present some descriptive statistics about the analysed authors and their characteristics. Next, we will show the results of two regression analyses. The first model displays the factors influencing mobility (Table 4) and the second model indicates the impact of mobility on scholars' career (Table 5).

Also, we will utilize a propensity score matching (PSM), for causal inference, to examine how mobility influences the scientific outcome. At the end, with comparing the centrality measures of mobile and non-mobile authors in the co-authorship networks, we will try to explain their position and role in these scientific networks.

### 4.1. Descriptive statistics

Table 3 shows the number of included male and female researchers in this study and the proportion of international mobility per discipline. Comparing the proportion of mobile to non-mobile researchers for each gender shows that women have been less mobile (37%) than men (44%) overall.

Figure 1 represents the distribution of mobile researchers based on their mobility score. The mobility score has a range from 1 to 58 and it has a skewed distribution. Most mobile researchers (~42%) have only one movement over their career life.

Table 3 Number of analysed authors and proportion of mobile and non-mobile among gender and disciplines. The percentages identify the proportion of non-mobile or mobile researchers to all researchers.

|  |  | Non mobile, Mobility Score=0 (percentage) | Mobile, mobility score>=1 (percentage) | Total |
|---|---|---|---|---|
| **Health Sciences** | Women | 83,107 (**65%**) | 43,916 (**35%**) | 127,023 |
|  | Men | 161,941 (**58%**) | 119,508 (**42%**) | 281,449 |
|  | Total | 245,048 (**60%**) | 163,424 (**40%**) | 408,472 |
| **Life Sciences** | Women | 55,132 (**61%**) | 34,915 (**39%**) | 90,047 |
|  | Men | 100,478 (**55%**) | 83,067 (**45%**) | 183,545 |
|  | Total | 155,610 (**57%**) | 117,982 (**43%**) | 273,592 |
| **Physical Sciences** | Women | 48,396 (**60%**) | 32,634 (**40%**) | 81,030 |
|  | Men | 187,143 (**54%**) | 157,812 (**46%**) | 344,955 |
|  | Total | 235,539 (**56%**) | 190,447 (**44%**) | 425,985 |
| **Social Sciences** | Women | 16,288 (**72%**) | 6,364 (**28%**) | 22,652 |
|  | Men | 34,681 (**65%**) | 18,973 (**35%**) | 53,654 |
|  | Total | 50,969 (**67%**) | 25,337 (**33%**) | 76,306 |
| **Total** | Women | 202,923 (**63%**) | 117,829 (**37%**) | 320,752 |
|  | Men | 484,243 (**56%**) | 379,360 (**44%**) | 863,603 |
|  | Total | 687,166 (**58%**) | 497,189 (**42%**) | 1,184,355 |

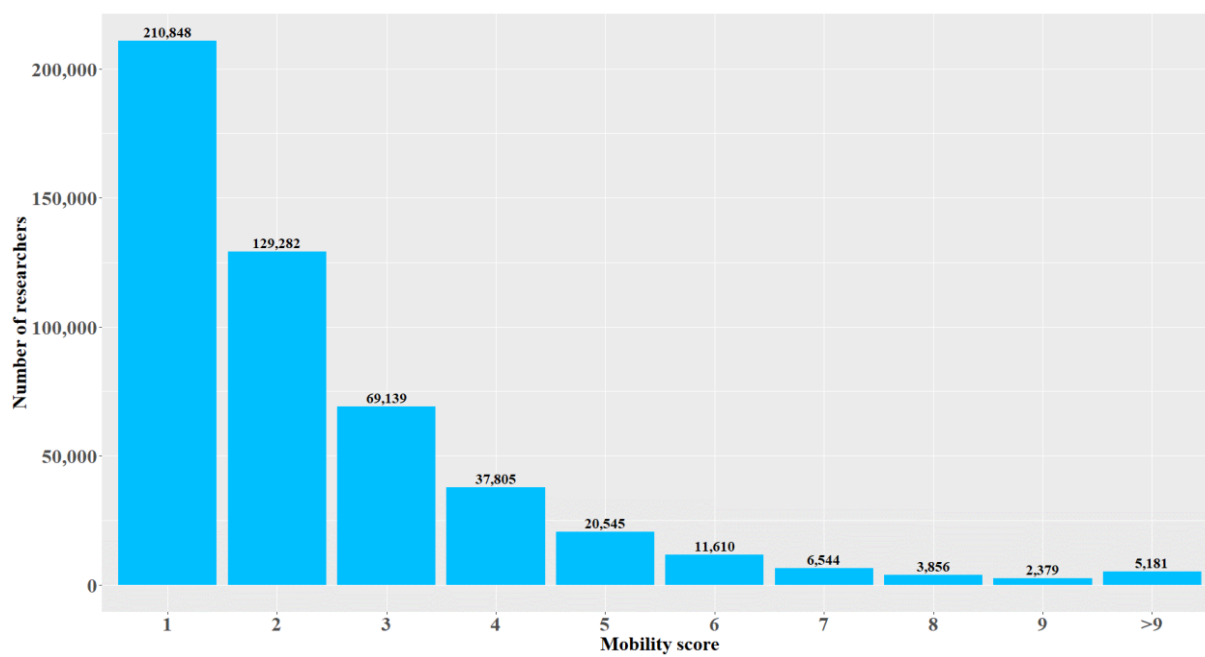

Figure 1 Distribution of mobile researchers based on their mobility score. The biggest group is for researchers with a mobility score of one and the number of researchers decreases for higher mobility scores.

*4.2. Mobility differences across countries, disciplines, and career stages*

Figure 2 (a) shows that in all scientific fields, women are underrepresented in international mobility, especially in Physical Sciences women have least participation. This agrees with the result of study by (Bhandari, 2017) and can be the result of gender inequality in this field ( (Wang & Degol, 2017); (Miyake et al., 2010)). From Figure 2 (b) we observe that in all fields the proportion of internationally mobile female to all female researchers is less than for male researchers. Physical Sciences and Life Sciences have the most mobile researchers for both genders. This complements the results of study by Aman (2020), which found knowledge transmitting by researchers in these fields are relatively high. The results in this figure agree with the results of prior studies (Bauder, 2020; Jöns, 2011; Leemann, 2010; Ryazanova & McNamara, 2019).

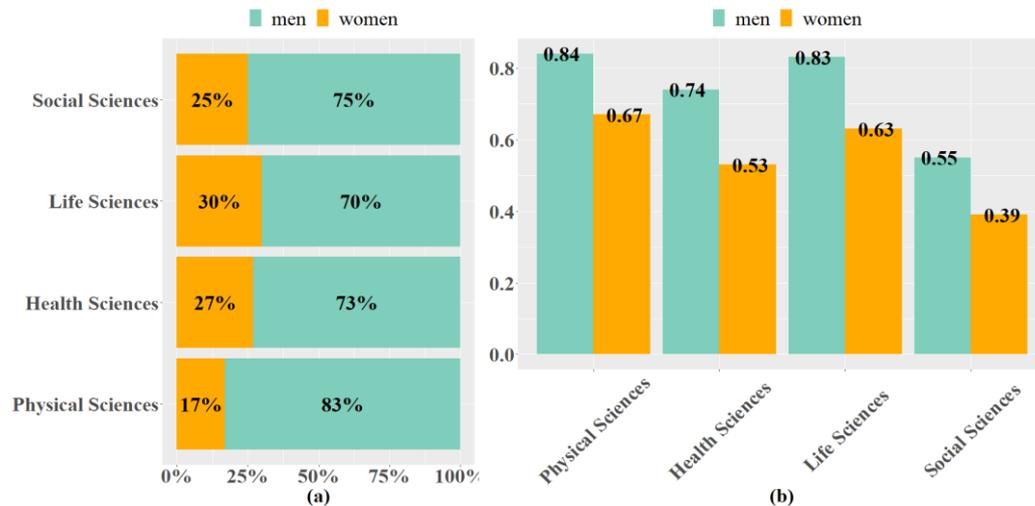

Figure 2 (a) The share of male and female mobile researchers across fields, Physical Sciences has the lowest percentage of mobile women among all fields. (b) Ratio of male/female mobile researchers to non-mobile male/female researchers across fields, Physical Sciences and Social Sciences have the highest and lowest participation in mobility for both genders, respectively.

Figure 3 displays 20 top origin countries of researchers with the number and percentage of mobile researchers (Figure 3 (a)) and the share of women among mobile researchers (Figure 3(b)). Figure 3 (a) reveals the lowest and highest percentage of mobile researchers for Turkey and Switzerland, respectively. From Figure 3(b), we observe that women from Japan have the lowest proportion in mobility among all countries. Interestingly Brazil, one of the BRICS countries, has the highest ratio of female mobile researchers. To show what bias can yield the gender detection method we build another dataset and apply all filters mentioned in the section 'Data and methods' except filtering for gender status. This dataset involves 1,878,545 authors. We compare Figure 3 (a) with the result for this dataset and present it in Appendix A.

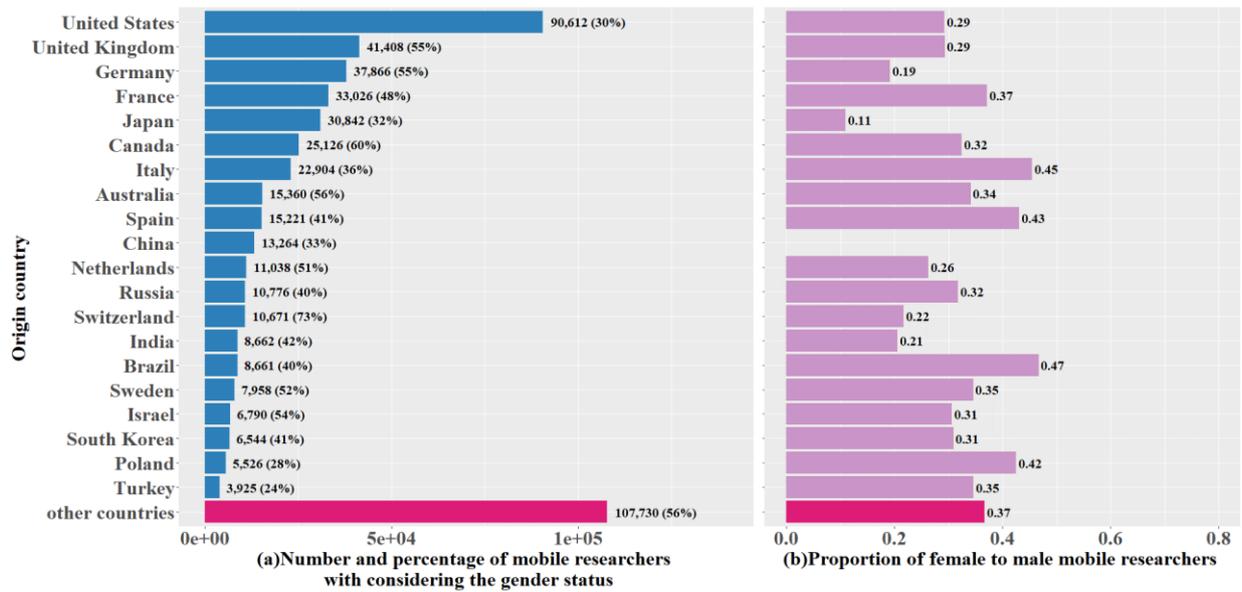

Figure 3 (a) distribution of mobile researchers by origin country grouped by 20 top countries with the largest number of mobile researchers and remained countries. The numbers show the number of mobile researchers and percentages in parentheses represent the percentage of mobile to all researchers (b) The proportion of female to male mobile researchers among countries, for all countries, the proportion is less than one and it means in general women participate in international mobility less than men. Note: given the fact that gender detection is very weak for Chinese names and we have tried to eliminate those names as much as possible from the raw data, we present no proportion for this country in part (b).

We calculate the mean mobility score of mobile researchers for both genders at each career stage. Figure 4 (a) shows the proportion of mobile men and women is stable through career stages and just one percent point decrease for men at mid and late career stages. Also, Figure 4 (b) indicates the highest mobility score in late-career stage for both genders, which disagrees with previous work by Leemann (2010) who found that probability to go abroad reduces by each age year for both men and women which is related to having family and children. After a decrease in movement in the mid-career stage for men, their mobility score grows again in the late career stages.

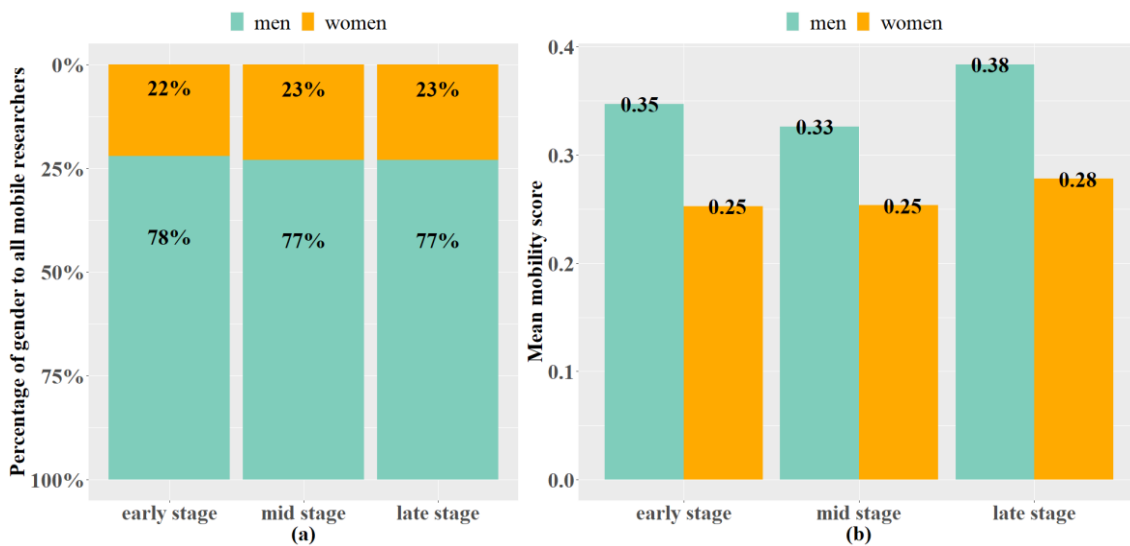

Figure 4 (a) Percentage of mobile men and women to all mobile researchers in three career stages. (b) Average mobility score of men and women in different career stages, the percentage numbers are the percentage change of mobility score from early to current career stage.

To display the flow of international mobility, we counted each movement among all authors and their career stages from one country to another. Figure 5 displays the aggregated results of the flow of mobility between continents and across career stages. These results show a higher tendency to move to "Europe & Central Asia" and "North America" in the early stage which slowly inclines to other regions in the late stage.

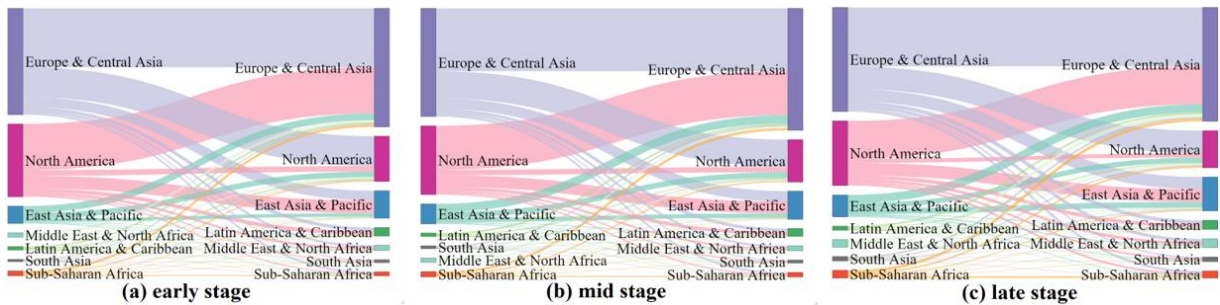

Figure 5 Movement between and within world regions. Each movement is a changing of the country affiliation for an author at a particular career stage. The right and left sides in each stage are regions before and after movement, respectively. Most movements happen between Europe & Central Asia for all career stages, but the tendency to move to this region has decreased in the late stage. Also, countries in this region are the most popular destination for authors from other regions. Movement from and to two regions "Middle East & North Africa" and Sub-Saharan Africa" have increased in the late career stage.

### 4.3. factors associated with mobility

In order to understand which socio-demographic characteristics associate with mobility and how mobility correlates with scholars' academic performance, we utilized two regression models. The first model shows the factors that relate to mobility at mid and later career stages considering the history of researchers at earlier career stages. Poisson regression is used for this model. Table 4 reports exponentiated coefficient, **exp(β)** of independent variables. In this model, values greater (less) than unity indicate positive (negative) correlation with dependent variable. To avoid confounding effects in this regression, we select only those researchers at career stage $s_i$ who were non-mobile at the past career stages. Thus, we can observe how those independent variables are related to mobility of researchers. Results show that having international co-authors at the previous career stage is the most significant factor in increasing the probability of international mobility. This complements the research by Bauder (2020) showing that international social capital facilitates international mobility. Besides, we observe that at the earlier career stages, the tendency for mobility in Social Sciences is less than in other fields, but social scientists are most likely to be mobile at the late-career stage. Researchers from the North America region have the highest GDP per capita and lowest probability of mobility across regions, respectively. Among all regions, Sub-Saharan Africa with a relative low GDP per capita has the most engaged researchers in mobility. Also, Chinchilla-Rodríguez, Liu and Bu (2021) reported the highest participation rate of international collaboration for this region. Interaction between gender and two other independent variables in this table, indicates the extent of gender inequality in mobility for different regions or scientific fields. For example, regarding the interaction between gender and region of origin, females from Sub-Saharan Africa are more likely to be mobile than those from other regions. Given the low value of $R^2$, our results suggest that in general it is not easy to predict the determinants of mobility. We denote that the very low p-value in regression results can be affected by the size of the sample. Hence, its representativeness for significance of statistical results may suffer in large-N settings. Therefore, only relying on low p-value is not sufficient to support the hypotheses. To reduce the p-value problem Lin et al. (2013) and Khalilzadeh and Tasci (2017) suggested some solutions (e.g., presenting effect size, reporting confidence intervals and using charts).

According to the proposed recommendations by Lin et al. (2013), we report the coefficients and their confidence intervals for Table 4 and Table 5 in the Appendix B.

Table 4 The results of Poisson regression. The dependent variable is the mobility score at career stage $s_i$ where i=2,3. The outputs are Odd ratio, $(\exp(\beta))$. $(1-\exp(\beta))$ shows the percentage change of dependent variable per unit increase in an independent variable, therefore numbers greater/less than one indicate a positive/negative correlation between variables. For interpreting the interaction between variables, we should multiply odd ratio of the related interaction to the odd ratio of both variables. For example, the odds for "Latin America & Caribbean and female" equals to 0.996*1.56*0.75=1.16 and the value more than one means that the likelihood of mobility for female researchers from Latin America & Caribbean is higher than males from North America.

|  | $\exp(\beta)$ at $s_1$ | $\exp(\beta)$ at $s_2$ | $\exp(\beta)$ at $s_3$ |
|---|---|---|---|
| Intercept | 0.26***(-280) | 0.02***(-236) | 0.06***(-247.7) |
| **Independent variables** | | | |
| Having international co-author at career stage $s_i - 1$ | ------ | 7.9***(137.3) | 3.53***(130) |
| *Gender:* | | | |
|    Male | Reference | Reference | Reference |
|    Female | 0.75***(-25.2) | 0.84***(-35.6) | 0.81***(-13.3) |
| *Region of origin country (Average GDP per Capita):* | | | |
|    North America (43,207) | Reference | Reference | Reference |
|    Latin America & Caribbean (13,110) | 1.56***(43.7) | 1.46***(28.01) | 1.33***(16.6) |
|    Europa & Central Asia (30,223) | 1.77***(119.1) | 1.50***(68.5) | 1.26***(31.5) |
|    Sub-Saharan Africa (6,197) | 2.00***(37.5) | 2.17***(36.12) | 2.35***(30.7) |
|    Middle East & North Africa (21,981) | 1.57***(38.8) | 1.54***(28.55) | 1.53***(23.4) |
|    South Asia (3,390) | 1.20***(12.9) | 1.74***(36.1) | 1.49***(19.8) |
|    East Asia & Pacific (26,783) | 1.11***(16.4) | 1.32***(44) | 0.90***(-10.0) |
| *Interaction between Region of origin country and Gender:* | | | |
|    North America and male | Reference | Reference | Reference |
|    Latin America & Caribbean and female | 0.87***(-6.9) | 0.93** (-2.6) | 0.89***(-4.2) |
|    Europa & Central Asia and female | 0.996 (-0.37) | 1.01 (0.7) | 0.87***(-8.8) |
|    Sub-Saharan Africa and female | 1.21***(4.60) | 1.11* (2.06) | 1.08 (1.42) |
|    Middle East & North Africa and female | 0.89***(-3.9) | 0.99 (-0.37) | 0.86***(-3.7) |
|    South Asia and female | 1.04 (1.14) | 1.07 . (1.7) | 1.00 (0.1) |
|    East Asia & Pacific and female | 1.32***(20.1) | 1.27***(13.4) | 1.26***(10.84) |
| *Field: Physical Sciences* | Reference | Reference | Reference |
|    Life Sciences | 0.93***(-15.6) | 1.07***(11.75) | 0.83***(-21.7) |
|    Health Sciences | 0.93***(-18.1) | 0.94***(-11.48) | 1.01 (1.5) |
|    Social Sciences | 0.47***(-68.0) | 0.87***(-12) | 1.45***(33.19) |
| *Interaction between Field and Gender:* | | | |
|    Physical Sciences and male | Reference | Reference | Reference |
|    Life Sciences and female | 0.86***(-15.4) | 0.88** (-2.9) | 0.97 . (5.2) |
|    Health Sciences and female | 0.998 (-0.16) | 0.96***(-8.6) | 1.08***(-1.74) |
|    Social Sciences and female | 0.96 . (-1.9) | 0.90***(-3.96) | 0.96 . (-1.74) |
| **Pseudo $R^2$** | 0.03 | 0.07 | 0.05 |
| **N** | 1,183,662 | 919,692 | 784,857 |

Significant codes: . p<0.1, * p < 0.05, ** p < 0.01, *** p < 0.001.
z-values of coefficients in parentheses

### 4.4. Impact of mobility on scholars' career

The second model demonstrates the relationship between mobility and co-author diversity, productivity, and citation. We chose ordinary least squares (OLS) regression for this purpose.

PPY, CPP and COPP are dependent variables in this model. To reduce the residual standard error of results, we used the log transformation of dependent variables.

Table 5 shows the regression results for men and women. Again, the results show exponentiated coefficient, exp(β) of independent variables. We observe a similar effect of mobility for both genders. The results show that mobility has better outcomes in terms of PPY, CPP for all mobile researchers and greater COPP for those who start mobility at the late career stage.

Table 5. OLS regression to estimate PPY, CPP and COPP. Dependent variables are log-transformed, therefore exponentiated coefficient of independent variables are presented. (1-exp(ß)) shows the percentage change of dependent variable per unit increase in an independent variable, therefore numbers greater/less than one indicate a positive/negative correlation between variables.

|  | PPY | | CPP | | COPP | |
|---|---|---|---|---|---|---|
|  | Men | Women | Men | Women | Men | Women |
|  | exp($\beta$) | exp($\beta$) | exp($\beta$) | exp($\beta$) | exp($\beta$) | exp($\beta$) |
| Intercept | 1.43***(223.1) | 1.44***(125.7) | 7.57***(1310) | 9.68***(846.5) | 1.68***324.4) | 1.77***(193.3) |
| **Independent variables:** | | | | | | |
| *Mobility score* | 1.15***(210.4) | 1.15***(98.8) | 1.06***(85.5) | 1.05***(37.3) | 1.05***(70.1) | 1.06***(38.03) |
| *Field*: | | | | | | |
|   Physical Sciences | Reference | Reference | Reference | Reference | Reference | Reference |
|   Health Sciences | 0.99 (-0.45) | 0.87***(-40.7) | 1.5***(210.7) | 1.47***(122.7) | 2.24***(388.8) | 2.25***(231.3) |
|   Life Sciences | 0.87***(-57.1) | 0.74***(-80.8) | 2.2***(351) | 2.09***(217.0) | 1.74***(236.7) | 1.77***(151.1) |
|   Social Sciences | 0.59***(-136.9) | 0.59***(-89.1) | 0.84***(-46.2) | 0.92***(-16.5) | 0.45***(-201.31) | 0.55***(-100.25) |
| *Career stage of first mobility:* | | | | | | |
|   Non-mobile | Reference | Reference | Reference | Reference | Reference | Reference |
|   Early stage | 1.37***(105.7) | 1.24***(42.7) | 1.18***(58.9) | 1.13***(26.7) | 0.92***(-26.1) | 0.96***(-8.33) |
|   Mid-stage | 1.33***(89.6) | 1.22***(38.6) | 1.27***(78.0) | 1.19***(35.7) | 0.98***(-6.2) | 0.99*** (-2.5) |
|   Late stage | 1.41***(100.9) | 1.36***(56.2) | 1.26***(71.4) | 1.18***(33.3) | 1.04***(11.7) | 1.05*(8.4) |
| R-square | 0.20 | 0.16 | 0.18 | 0.17 | 0.24 | 0.24 |
| Residual standard error | 0.82 | 0.77 | 0.78 | 0.70 | 0.81 | 0.78 |
| N | 863,595 | 320,744 | 860,837 | 320,114 | 859,568 | 319,782 |
| Significant codes <* $p < 0.05$, ** $p < 0.01$, *** $p < 0.001$. z-values in parentheses | | | | | | |

To estimate the effect of mobility by accounting for the covariates, we use propensity score matching, which calculates the causal effects of the treatment (mobility). To this end, we selected the 1:1 nearest-neighbour matching method to pair mobile with non-mobile authors that share similar characteristics and close research attributes at the stage before mobility. Then we compare the scientific outcomes regarding PPY, CPP, and COPP for these two groups. We select nine covariances for matching. To control the effect of the starting stage of mobility, we divided mobile researchers according to the career stage in which they start to be mobile and paired them separately with non-mobile researchers.

*Table 6* shows these covariates and Standardized Mean Difference (SMD) to assess the covariance balance before and after matching. Stuart et al. (2013) recommended threshold equals to 0.1 for asserting covariance balance between two groups. All SMDs are less than the threshold after matching and it shows that all covariates were balanced. Table 7 reports the result of t-tests which reveals higher PPY and CPP for mobile researchers.

Table 6. The Balance report before and after matching treatment (mobile) and control (non-mobile) groups; mobile researchers with different stage of stating mobility have been paired separately with non-mobile researchers through nine covariances.

| First mobility stage | $s_1$ | | $s_2$ | | $s_3$ | |
|---|---|---|---|---|---|---|
| | SMD Before matching | SMD After matching | SMD Before matching | SMD After matching | SMD Before matching | SMD After matching |
| **Covariance:** | | | | | | |
| Gender | 0.16 | 0.008 | 0.10 | 0.012 | 0.1 | 0.006 |
| Career age | 0.39 | 0.036 | 0.28 | 0.007 | 0.50 | 0.008 |
| Region | 0.33 | 0.041 | 0.22 | 0.019 | 0.16 | 0.013 |
| GPD per capita of the first affiliation country | 0.05 | 0.064 | 0.07 | 0.016 | 0.02 | 0.020 |
| Field | 0.20 | 0.021 | 0.13 | 0.005 | 0.08 | 0.018 |
| Having an international co-author | 0.65 | 0.001 | 0.6 | <0.001 | 0.58 | 0.004 |
| CPP at the previous stage and before mobility | - | - | 0.14 | 0.001 | 13.6 | 0.016 |
| PPY at the previous stage and before mobility | - | - | 0.36 | 0.003 | 1.85 | 0.006 |
| COPP at the previous stage and before mobility | - | - | 0.17 | 0.006 | 0.17 | 0.02 |
| **Sample sizes** | | | | | | |
| Control (non-mobile) | 679,456 | 260,169 | 679,456 | 133,420 | 637,446 | 90,922 |
| Treated (mobile) | 262,294 | 260,169 | 133,829 | 133,420 | 90,977 | 90,922 |
| Paired matched | - | 260,169 | - | 133,420 | - | 90,922 |

Table 7 The result of paired-samples t-test. Positive Mean diff. shows a higher outcome (PPY, CPP and COPP) for mobile researchers. PPY and CPP for mobile researchers are higher than non-mobile researchers regardless of the stage of starting mobility. Mobile researchers have the better outcome in terms of COPP than non-mobile researchers, only if they start mobility at the late career stage.

| First mobility stage | $s_1$ | | | $s_2$ | | | $s_3$ | | |
|---|---|---|---|---|---|---|---|---|---|
| | Mean Diff. | SE | t (p-value) | Mean Diff. | SE | t (p-value) | Mean Diff. | SE | t (p-value) |
| PPY | 1.93 | 0.01 | 198.1 (***) | 0.79 | 0.01 | 83.5 (***) | 0.71 | 0.01 | 69.9 (***) |
| CPP | 2.91 | 0.01 | 94.4 (***) | 2.5 | 0.04 | 54.6 (***) | 1.8 | 0.05 | 32.9 (***) |
| COPP | -0.68 | 0.06 | -11.5 (***) | -0.23 | 0.07 | -3.1 (***) | 0.13 | 0.1 | 1.2 (***) |
| n | 260,169 | | | 133,420 | | | 90,922 | | |
| Significant codes: p<0.1,* p < 0.05, ** p < 0.01, *** p < 0.001. | | | | | | | | | |

### 4.5. Position of mobile researchers in the co-authorship networks

In this section, we will compare the centrality scores of mobile and non-mobile researchers in the co-authorship network. Because the structure of co-authorship networks is dynamic (it changes yearly) and the position of authors in the network may change by their career development, we compare the centrality scores of researchers at the same career stage and for each year separately. To this end, we chose the researchers who started publishing at the same year and tracked their centrality scores for the future years. We represent the results of analyses for authors who started publishing in the arbitrary year (2000), but it is generalizable to other years.

We divided the data into non-mobile and mobile groups. From the mobile group, we selected authors who have mobility at the early or mid-career stage. To calculate the yearly centrality score of each group, first, we computed the average centrality of each author across networks which she/he belongs to them and considered it as her/his centrality score. Then we calculated the average score among researchers.

Figure 6 shows yearly centrality scores of selected researchers. For mobile researchers, we observe a growing trend in degree and coreness by increasing the career age. These two observations suggest that not only mobility increases the social capital of the scholars (degree), but also it helps to position the scholars into the core of the community which would allow them to get better access to highly influential people (Kitsak et al., 2010).

In contrast, the lower clustering coefficient of mobile researchers indicates that by increasing their degree, they diversify their co-authors and belong to various communities and thus they act more as bridges between the communities.

We note that the relationship between mobility and scholars' careers that we observed in Table 5 could be to some extent explained by the collaboration networks (Jadidi et al., 2018) as we see in Figure 6. Given the significant influence of mobility on centrality and position of researchers in their collaboration network, this also confirms the importance of mobility in advancing one's career.

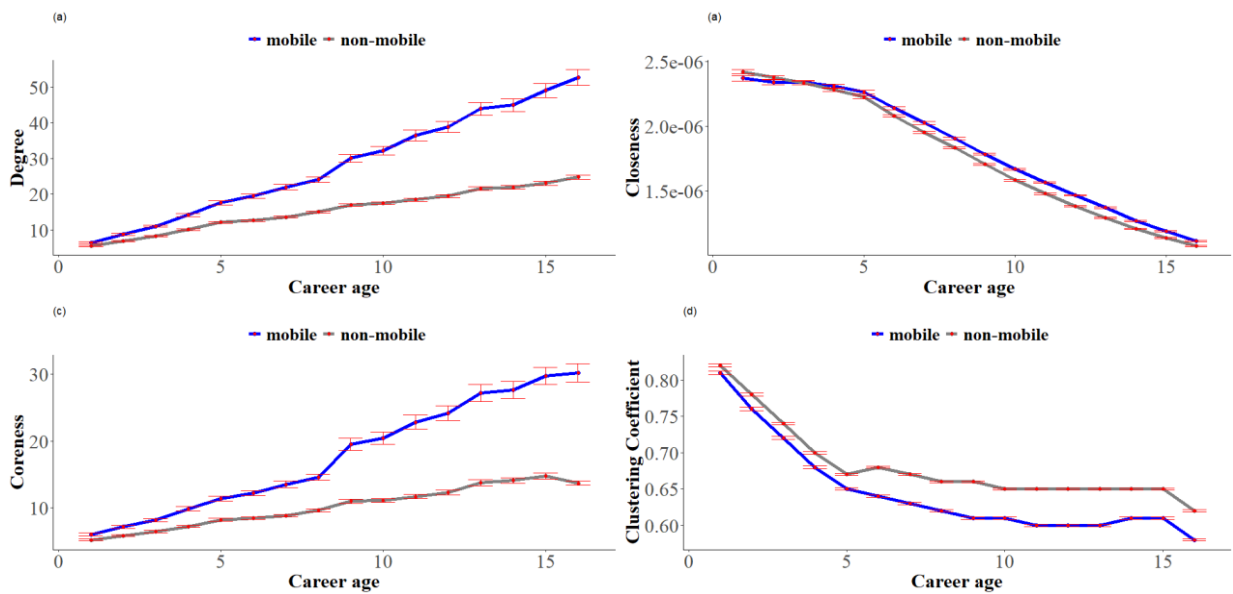

Figure 6. Centrality measures of mobile and non-mobile researchers. Number of observations are 18,734 and 29,122 for mobile and non-mobile groups respectively. Error bars show standard errors. Career age has a range of 1 to 16 (first publication year and 15 years after that).

## 5. Discussion and conclusion

In this paper, we have identified and explored many facets of academic mobility of males and females that relate to 1) mobility differences across nationalities, disciplines, and career stages, 2) the international mobility flow 3) the factors that associate with mobility, 4) relationship between mobility and scholars' publications and citations, 5) different position of mobile and non-mobile scholars in their collaboration network.

Our findings regarding the first research question revealed that in the world of academia, international mobility of women is less than men, which is consistent with the past studies (Jöns, 2011; Leemann, 2010; Myers & Griffin, 2019; Toader & Dahinden, 2018). We observe that the mobility score of both genders has the lower level in the early and mid-career stage, which can be related to the time that families should focus more on preschool children than going abroad. The results point that gender inequality in international mobility exists for all scientific fields and women are underrepresented particularly in Physical Sciences. Physical Sciences and Life Sciences have the most mobile researchers for both genders, which agrees with the findings by Subbotin and Aref (2021). Our findings show that mobility of researchers decreases with increasing the career age for both genders which is in line with the results of the study by Leemann (2010). The regression result of the first model pointed to the importance of having international collaboration for mobility in future. Also, we observed various tendencies for mobility across geographic regions, scientific fields, and career stages. Researchers from the Sub-Saharan Africa region with relatively low-income levels are most likely to move. Females from this region have the highest probability to be mobile as well. It seems that getting better funding opportunities is a motivation for researchers to go abroad, as Hunter et al. (2009) found that top scholars head to countries with high R&D spending levels. From the results, authors in the Social Sciences have the least and most probability for mobility in the mid and late-career stages, respectively. Receiving postdoctoral positions in this field might be hard for international researchers and a reason for low participation in the mid-career stage. Perhaps these researchers should gain academic experiences at the earlier career stage in their own country to increase their chance of receiving an international position.

Regarding the second research question, the results of our second regression analysis demonstrated the relationship between mobility with the performance, success, and number of unique co-authors of researchers. We compared the outcomes of mobility for men and women and found no clear difference between them. We observed that although mobility improves the performance and success of researchers regardless of the stage of starting mobility. We used PSM to draw the causal reference and minimize the confounding bias in examining the effect of mobility. By matching mobile with non-mobile researchers with the similar characteristics, we compared their scientific outcome and found the positive impact of mobility on scientific performance and success.

Finally, the co-authorship networks of mobile and non-mobile researchers reveals that mobile researchers have a more diverse social capital and better access to influential scholars in their network compared to non-mobile researchers.

The results carry potential insights for policy-makers concerned with the issues and inequalities in this area to provide fair opportunities at the proper phase of the academic career for all researchers who desire to communicate and collaborate with their international colleagues.

## 6. Limitations and future work

This study has some limitations that should be noted. We used bibliometric data to indicate the mobility and career age of the authors. We used Scopus author ID to associate authors with their publications. Although this author ID system has high precision to assign the set of articles to a particular author ID, an author ID may not cover the complete article set of an author and result in multiple author IDs for one author in Scopus (Moed et al., 2013). It causes problems in tracking the affiliation of authors whose works have been split into multiple author IDs and may underestimate the mobility of authors. Besides, for publications that aren't indexed in Scopus, this approach can lead to errors in specifying the country of origin, mobility score, and researcher's career stage. Next, we have assumed a fixed period of early and mid-career stages for all authors. These can vary depending on the study field, career interruption, or type of affiliate organization. Although a comprehensive discussion of the few related works is included in Section 'Data & Methods', which argues in favour of fixed periods, future work would need to consider the robustness of this approach. Also, we didn't distinguish between academic researchers and those outsides academia (industry or government researchers) who have varying average number of publications and career advancement. In addition, this approach cannot indicate temporary mobilities, such as research visits, that the host countries are not considered as the author's affiliation. The collaboration pattern of these may differ from other mobile researchers. For example, they are less likely to lose co-authors from their country of origin. Moreover, our analyses don't contain authors for whom we could not detect their gender status, because the applied method has some weaknesses for common names especially Chinese names.

In this study we built the collaboration networks for each discipline separately. This might affect the network position of scholars who work on interdisciplinary topics. In the future, more attention is needed to study the influence of mobility and collaboration networks on researchers with interdisciplinary backgrounds. Also, we didn't control for the size of the network or communities that researchers belong to (one can start academic life in an organisation with high density and interaction between authors, others in a much more isolated one) and that may impact the future career perspectives.

In this paper, we considered all publications without regarding the position of authors in the paper. By this means we make statements about all effects of collaboration not only 'prestigious effects' that lead to becoming a first author or a last author. In future, it would be interesting to consider the influence of mobility on the authorship position.

In this study we didn't consider destination countries of researchers and different kinds of mobility (e. g. immigrants, returnees). Including the different individual-level mobility introduced by Robinson-Garcia et al. (2019) in analyses leads to more comprehensive results and give us a better knowledge about the motivation of movement, advantages, and disadvantages of mobility for the origin and destination countries.

We looked at the mobility of authors at the country level. Investigating the mobility in scientific fields and its relation to geographic mobility will give us a better understanding of the impact of mobility on knowledge transfer between scientific fields. Also, it would be interesting to analyse the co-authorship networks among different kinds of mobile researchers to discover the differences in collaboration pattern and their impact on other researchers.

During the COVID-19 pandemic, international mobility has been limited and many countries have reduced international academic positions. In the meantime, telecollaboration has become more popular, which makes international collaboration easier. Investigating the impact of the pandemic on the scholars' career and comparing collaboration patterns before and after the pandemic, will give us an understanding of the importance of physical mobility. Perhaps virtual mobility can be an alternative for scientists.

**Competing interests**

 We declare we have no competing interests.

**Acknowledgments**

This work is supported by BMBF project OASE, grant number 01PU17005A. We acknowledge the support of the German Competence Center for Bibliometrics (grant: 01PQ17001). We are thankful to Nina Smirnova for her aid with analysing data. We thank Dr. Matthias Raddant and Dr. Anne-Kathrin Stroppe for helpful comments.

# Appendix A

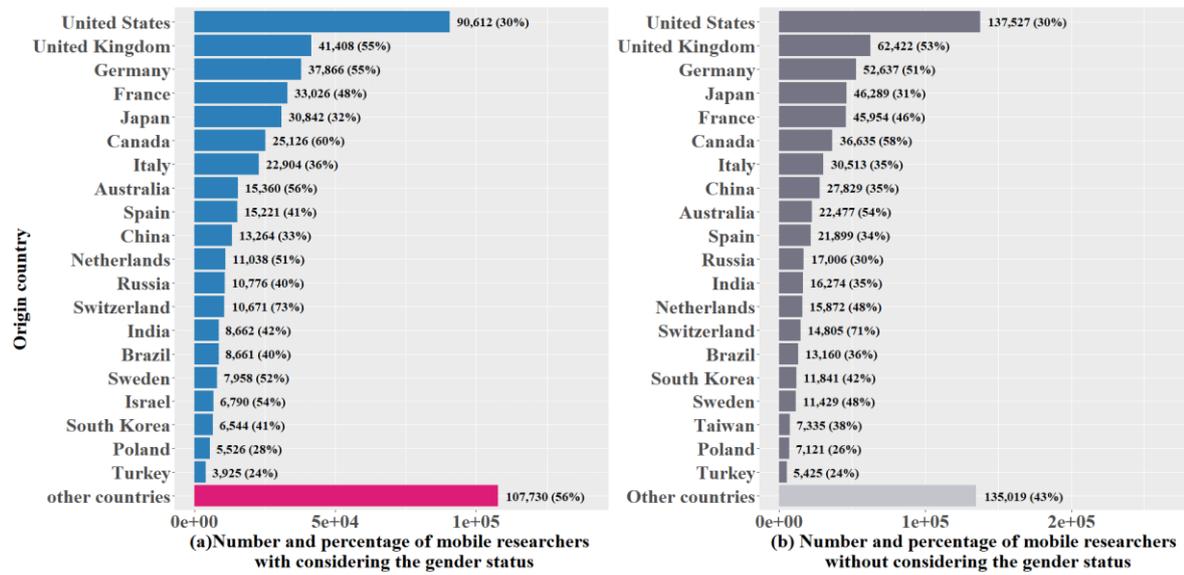

Figure A. Comparison of distribution of mobile researchers by origin country for two samples: (a) The sample with gender status and (b) The sample without gender status. Some Asian countries such as China, South Korea and India have a lower ranking in the sample with gender status.

## Appendix B

Table B1 Coefficients and confidence intervals for the Poisson regression presented in Table *4*.

| | $s_1$ | | | $s_2$ | | | $s_3$ | | |
|---|---|---|---|---|---|---|---|---|---|
| | β* | 2.5% CI** | 97.5% CI*** | β | 2.5% CI | 97.5% CI | β | 2.5% CI | 97.5% CI |
| Intercept | -1.32 | -1,33 | -1,32 | -3.7 | -3.77 | -3.71 | -2.8 | -2.86 | -2.81 |
| **independent variables** | | | | | | | | | |
| *Gender:* | | | | | | | | | |
| male | | Reference | | | Reference | | | Reference | |
| female | -0.28 | -0.30 | -0.21 | -0.17 | -0.21 | -0.14 | -0.21 | -0.25 | -0.18 |
| Having international co-author at career stage $s_i - 1$ | ------ | | | 2.1 | 2.04 | 2.1 | 1.26 | 1.24 | 1.28 |
| *Region of origin country (Average GDP per Capita):* | | | | | | | | | |
| North America (43,207) | | Reference | | | Reference | | | Reference | |
| Latin America & Caribbean (13,110) | 0.44 | 0.43 | 0.47 | 0.37 | 0.35 | 0.40 | 0.28 | 0.24 | 0.31 |
| Europa & Central Asia (30,223) | 0.57 | 0.56 | 0.58 | 0.40 | 0.39 | 0.41 | 0.23 | 0.21 | 0.24 |
| Sub-Saharan Africa (6,197) | 0.69 | 0.65 | 0.73 | 0.77 | 0.72 | 0.82 | 0.85 | 0.80 | 0.91 |
| Middle East & North Africa (21,981) | 0.45 | 0.42 | 0.47 | 0.43 | 0.40 | 0.46 | 0.42 | 0.39 | 0.46 |
| South Asia (3,390) | 0.18 | 0.15 | 0.21 | 0.55 | 0.52 | 0.59 | 0.40 | 0.36 | 0.44 |
| East Asia & Pacific (26,783) | 0.10 | 0.10 | 0.12 | 0.28 | 0.26 | 0.29 | -0.1 | -0.12 | -0.08 |
| *Interaction between Region of origin country and Gender:* | | | | | | | | | |
| North America and male | | Reference | | | Reference | | | Reference | |
| Latin America & Caribbean and female | -0.14 | -0.18 | -0.01 | -0.07 | -0.12 | -0.02 | -0.12 | -0.19 | -0.07 |
| Europa & Central Asia and female | -0.003 | -0.02 | 0.01 | 0.00 | -0.02 | 0.03 | -0.13 | -0.16 | -0.10 |
| Sub-Saharan Africa and female | 0.18 | 0.10 | 0.26 | 0.10 | 0.00 | 0.2 | 0.08 | -0.03 | 0.19 |
| Middle East & North Africa and female | -0.11 | -0.16 | -0.06 | -0.01 | -0.08 | 0.05 | -0.14 | -0.22 | -0.07 |
| South Asia and female | 0.04 | -0.03 | 0.1 | 0.07 | -0.01 | 0.15 | 0.004 | -0.09 | 0.09 |
| East Asia & Pacific and female | 0.28 | 0.25 | 0.30 | 0.24 | 0.20 | 0.27 | | 0.18 | 0.26 |
| *Field* : Physical Sciences | | Reference | | | Reference | | | Reference | |
| Life Sciences | -0.08 | -0.09 | -0.07 | 0.08 | 0.07 | 0.09 | -0.19 | -0.20 | -0.17 |
| Health Sciences | -0.08 | -0.08 | -0.07 | -0.03 | -0.05 | -0.02 | 0.01 | 0.00 | 0.02 |
| Social Sciences | -0.7 | -0.8 | -0.70 | -0.11 | -0.13 | -0.08 | 0.37 | 0.35 | 0.39 |
| *Interaction between Field and Gender:* | | | | | | | | | |
| Physical Sciences and male | | Reference | | | Reference | | | Reference | |
| Life Sciences and female | -0.001 | -0.02 | 0.17 | -0.04 | -0.07 | -0.01 | -0.03 | -0.07 | 0.003 |
| Health Sciences and female | -0.15 | -0.17 | -0.13 | -0.12 | -0.15 | -0.09 | 0.08 | 0.05 | 0.11 |
| Social Sciences and female | -0.04 | -0.1 | 0.0 | -0.10 | -0.15 | -0.05 | -0.04 | -0.08 | 0.00 |

*Coefficient
** Coefficient at confidence interval: 2.5%
***Coefficient at confidence interval: 97.5%

Table B2 Coefficients and confidence intervals for the OLS regression presented in Table 5.

|  | PPY | | | | | | CPP | | | | | | COPP | | | | | |
|---|---|---|---|---|---|---|---|---|---|---|---|---|---|---|---|---|---|---|
|  | Men | | | Women | | | Men | | | Women | | | Men | | | Women | | |
|  | β* | 2.5% CI** | 97.5% CI*** | β | 2.5% CI | 97.5% CI | β | 2.5% CI | 97.5% CI | β | 2.5% CI | 97.5% CI | β | 2.5% CI | 97.5% CI | β | 2.5% CI | 97.5% CI |
| Intercept | 0.36 | 0.36 | 0.36 | 0.37 | 0.36 | 0.37 | 2.02 | 2.02 | 2.03 | 2.27 | 2.26 | 2.27 | 0.52 | 0.52 | 0.52 | 0.57 | 0.57 | 0.58 |
| **independent variables:** | | | | | | | | | | | | | | | | | | |
| *mobility score* | 0.15 | 0.14 | 0.14 | 0.14 | 0.14 | 0.14 | 0.06 | 0.06 | 0.06 | 0.05 | 0.05 | 0.05 | 0.05 | 0.05 | 0.05 | 0.05 | 0.05 | 0.06 |
| *Field*: | | | | | | | | | | | | | | | | | | |
| Physical Sciences | Reference | | | Reference | | | Reference | | | Reference | | | Reference | | | Reference | | |
| Health Sciences | -0.001 | -0.005 | 0.003 | -0.14 | -0.15 | -0.13 | 0.42 | 0.41 | 0.42 | 0.39 | 0.38 | 0.39 | 0.81 | 0.80 | 0.81 | 0.80 | 0.8 | 0.81 |
| Life Sciences | -0.13 | -0.14 | -0.13 | -0.30 | -0.31 | -0.29 | 0.79 | 0.79 | 0.80 | 0.74 | 0.73 | 0.74 | 0.56 | 0.55 | 0.56 | 0.56 | 0.56 | 0.58 |
| Social Sciences | -0.52 | -0.53 | -0.51 | -0.51 | -0.53 | -0.50 | -0.17 | -0.18 | -0.16 | -0.09 | -0.1 | -0.07 | -0.78 | -0.79 | -0.78 | -0.6 | -0.61 | -0.58 |
| *Career stage of first mobility:* | | | | | | | | | | | | | | | | | | |
| Non-mobile | Reference | | | Reference | | | Reference | | | Reference | | | Reference | | | Reference | | |
| Early stage | 0.31 | 0.31 | 0.32 | 0.22 | 0.21 | 0.23 | 0.17 | 0.16 | 0.17 | 0.12 | 0.11 | 0.13 | -0.08 | -0.08 | -0.07 | 0.04 | -0.05 | -0.03 |
| Mid-stage | 0.28 | 0.28 | 0.29 | 0.20 | 0.19 | 0.21 | 0.24 | 0.23 | 0.24 | 0.17 | 0.16 | 0.18 | -0.04 | -0.03 | -0.05 | -0.01 | -0.002 | -0.003 |
| Late stage | 0.34 | 0.34 | 0.35 | 0.31 | 0.30 | 0.32 | 0.23 | 0.23 | 0.24 | 0.17 | 0.16 | 0.18 | 0.04 | 0.03 | 0.05 | 0.05 | 0.04 | 0.06 |

*Coefficient
** Coefficient at confidence interval: 2.5%
***Coefficient at confidence interval: 97.5%